# How Cooper pairs vanish approaching the Mott insulator in $Bi_2Sr_2CaCu_2O_{8+\delta}$

Y. Kohsaka[1,2], C. Taylor[1], P. Wahl[1], A. Schmidt[1], Jhinhwan Lee[1], K. Fujita[1,3], J. W. Alldredge[1], Jinho Lee[1,4], K. McElroy[5], H. Eisaki[6], S. Uchida[3], D.-H. Lee[7] & J. C. Davis[1,4]

*1LASSP, Department of Physics, Cornell University, Ithaca, New York 14853, USA.*

*2RIKEN, Wako, Saitama 351-0198, Japan.*

*3Department of Physics, University of Tokyo, Bunkyo-ku, Tokyo 113-0033, Japan.*

*4CMPMS Department, Brookhaven National Laboratory, Upton, New York 11973, USA.*

*5Department of Physics, University of Colorado, Boulder, Colorado 80301, USA.*

*6Institute of Advanced Industrial Science and Technology, Tsukuba, Ibaraki 305-8568, Japan.*

*7Department of Physics, University of California, Berkeley, California 94720, USA.*

**The antiferromagnetic ground state of copper oxide Mott insulators is achieved by localizing an electron at each copper atom in real space (r-space). Removing a small fraction of these electrons (hole doping) transforms this system into a superconducting fluid of delocalized Cooper pairs in momentum space (k-space). During this transformation, two distinctive classes of electronic excitations appear. At high energies, the enigmatic 'pseudogap' excitations are found, whereas, at lower energies, Bogoliubov quasi-particles—the excitations resulting from the breaking of Cooper pairs—should exist. To explore this transformation, and to identify the two excitation types, we have imaged the electronic structure of**

**$Bi_2Sr_2CaCu_2O_{8+\delta}$ in r-space and k-space simultaneously. We find that although the low energy excitations are indeed Bogoliubov quasi-particles, they occupy only a restricted region of k-space that shrinks rapidly with diminishing hole density. Concomitantly, spectral weight is transferred to higher energy r-space states that lack the characteristics of excitations from delocalized Cooper pairs. Instead, these states break translational and rotational symmetries locally at the atomic scale in an energy independent fashion. We demonstrate that these unusual r-space excitations are, in fact, the pseudogap states. Thus, as the Mott insulating state is approached by decreasing the hole density, the delocalized Cooper pairs vanish from k-space, to be replaced by locally translational- and rotational-symmetry-breaking pseudogap states in r-space.**

**Evolution from $CuO_2$ superconductor to Mott insulator**

Theoretical studies of the copper oxide electronic structure using the Hubbard model[1–13] suggest a complex phenomenology for the evolution from the localized r-space states of the Mott insulator to the delocalized k-space Cooper pairs of the superconductor. The unique capability of quasi-particle interference (QPI) imaging[14] to determine the electronic structure simultaneously in r-space and k-space makes it ideal for studying such effects. Here we use superconducting QPI[15–19] to study the evolution of electronic structure as hole density p tends towards zero in $Bi_2Sr_2CaCu_2O_{8+\delta}$. We find that the Bogoliubov quasi-particle (BQP) excitations are confined to a specific locus in k-space called the 'Bogoliubov arc'. The end points of this arc, where BQP interference disappears, lie near the diagonal lines connecting $\mathbf{k} = (0, \pm\pi/a_0)$ and $\mathbf{k} = (\pm\pi/a_0, 0)$ and occur at a weakly doping-dependent 'extinction energy' $E = \Delta_0$ ($a_0$ is the nearest neighbour Cu-Cu distance). This Bogoliubov arc, which represents regions of $\mathbf{k}$-space supporting delocalized Cooper pairs, shrinks rapidly as $p$ decreases.

For $E > \Delta_0$, the excited states appear dramatically different. They do not have the BQP interference phenomena expected of excitations from delocalized Cooper pairs. Instead, they appear to be quasi-localized in **r**-space. Also, energy-resolved tunnelling asymmetry[20], when corrected appropriately for nanoscale electronic disorder, reveals that the intensity of these **r**-space characteristics is most pronounced at the pseudogap energy. Thus the low-p pseudogap excitations locally break translational symmetry, and reduce the 90° rotational ($C_4$) symmetry of each four-Cu-atom plaquette to a 180° rotational symmetry ($C_2$) in Cu–O–Cu bond-centred patterns[20].

**Two classes of CuO$_2$ excited states at low hole density**

The hole-doped CuO$_2$ plane has a complex electronic phase diagram[21]. Its ground states include antiferromagnetism for $p < 2\%$–$5\%$, d-wave superconductivity for $5\%$–$10\% < p < 25\%$, and a metallic state for $p > 25\%$. Several poorly understood regions occur at finite temperatures in the non-superconducting state, most significantly the pseudogap regime[21]. Its spectrum of electronic excitations reveals the pseudogap phenomenology most transparently. Two distinct energy scales, associated with two distinct types of excited state dynamics, can be detected by numerous spectroscopies[22]. The two excitation energies diverge with diminishing p as shown, for example, in Fig. 1a (reproduced from ref. 22, the nomenclature of which we follow in this introduction).

Single-particle tunnelling spectroscopy[23] detects a particle–hole-symmetric excitation energy $E_{PG}$ that is indistinguishable in magnitude in the pseudogap and superconducting states[23] and which increases with decreasing $p$. A typical evolution of such spectra with decreasing $p$ is shown in Fig. 1b. Here the two classes of excitations are respectively spatially heterogeneous[24–26] excitations centred on $E_{PG}$ and spatially homogeneous[16,19,24–26] excitations surrounding $E = 0$; they exhibit increasing energy segregation as $p \to 0$. Andreev–St James tunnelling, which is sensitive to the phase-coherent superconducting state, also reveals two diverging excitation energy scales as $p$

→ 0: the first is the familiar high-energy scale $E_{PG}$ but the second, lower energy, scale $E_{SC}$ is viewed as the maximum energy of Cooper-pair binding[27]. Optical transient-grating spectroscopy[28] shows that the higher energy excitations near $E_{PG}$ propagate exceedingly slowly and without recombination to form Cooper pairs, whereas lower energy excitations surrounding the $E = 0$ nodes propagate and dynamically reform Cooper pairs as expected. Raman scattering spectroscopy[29] reveals the two distinct energy scales and quasi-particle dynamics in the nodal and antinodal regions of **k**-space, again diverging as $p \to 0$.

Finally, angle-resolved photoemission spectroscopy (ARPES) plays a central role in categorizing these two excitation types. In the pseudogap regime, ARPES reveals an excitation energy $E_{PG}$ in the antinodal region of momentum space near **k** ≈ (±π/$a_0$, 0) and **k** ≈ (0, ±π/$a_0$) that increases as $p \to 0$. $E_{PG}$ is independent of temperature, changing neither with greatly increased temperatures nor during the transition into the superconducting state at low temperature[22,30–34]. In contrast, the nodal region supports a 'Fermi arc'[30] of coherent excitations in the pseudogap state, upon which a momentum- and temperature-dependent gap opens only in the superconducting state[30,32–34]. A schematic representation of this situation (with $E_{PG}$ in blue and $E_{SC}$ in orange, following ref. 22) is shown in Fig. 1c. Notably, $E_{SC}$ links smoothly to the pre-existing pseudogap excitation energy $E_{PG}$ at their common point in **k**-space[30–34]. To understand the copper oxide superconductor-to-Mott insulator transition fully, it is thus critical to understand these two electronic excitation types along with their growing energetic segregation as $p \to 0$.

**Numerical studies of hole doped CuO$_2$ Mott insulators**

The hypothesis of a 'resonating valence-bond' state[35] has motivated the theoretical exploration of whether repulsive electron–electron interactions in the CuO$_2$ plane could provide the explanation for high-temperature superconductivity. In such proposals, when the antiferromagnetic Mott insulator is 'quantum melted' by hole doping, singlet electron-pairing correlations develop, with the pairing energy diminishing as p increases. However, such theoretical approaches have faced a number of challenges, including how to adequately represent the anisotropic electronic structure in **k**-space[29–34] and the quasi-localized[20,24–26] **r**-space states, which become increasingly important as $p \to 0$.

Recent theoretical advances in the numerical analysis of the Hubbard model[1–9] with associated analytical approximations[10–13] are highly relevant to these issues. They yield the following results, among others: the lightly hole-doped Mott insulator state has two excitation energy scales that diverge with decreasing $p$ (refs 1, 2, 7–9); the smaller scale represents the energy gap to excitations of delocalized Cooper pairs, whereas the larger scale is due to correlations and not a different order parameter[1–3,7–9]; a growing **k**-space anisotropy between these two types of excitation leads to an eventual break-up of the Fermi surface[1,4,5,7–11,13] as $p \to 0$; the Luttinger theorem relating the areas enclosed by zero-energy **k**-space contours to carrier density must thus be amended[1,9–13]; and the doped Mott insulator correlations produce an asymmetry of tunnelling probabilities for electron injection versus extraction[3,4,6,9,13] that diverges with decreasing $p$.

**Simultaneous r-space and k-space electronic structure determination**

To explore these predictions, the electronic structure must be determined simultaneously in **r**-space and **k**-space, as a function of decreasing $p$. Atomic-resolution spectroscopic-imaging scanning tunnelling microscopy in combination with BQP interference[15–19] is one of the few suitable approaches. This technique has a critical advantage in that it can detect the quantum coherence of **k**-space excitations directly from their interference patterns.

A distinctive form of QPI occurs in copper oxide superconductors because the Bogoliubov quasi-particle dispersion $E(\mathbf{k})$ has closed constant-energy contours surrounding the d-wave nodes as shown schematically by grey lines in Fig. 2a. Maxima in the joint-density-of-states occur at the eight tips of these closed contours, $\mathbf{k}_j(E)$, $j = 1, 2, \ldots, 8$, as shown by colour-coded symbols in Fig. 2a. For a given $E$, these $\mathbf{k}_j(E)$ occur at the minima (maxima) in the band structure of Bogoliubov excitations $\mathbf{k}_B(E)$ and at the normal-state Fermi surface. Elastic scattering between the $\mathbf{k}_j(E)$ is represented by the coloured arrows in Fig. 2a, b and produces $\mathbf{r}$-space interference patterns in the local density of states $N(\mathbf{r}, E)$. The resulting energy-dispersive $N(\mathbf{r}, E)$ modulations have 16 $\pm\mathbf{q}$ pairs of wavevectors, indicated by the colour-coded markers in Fig. 2b. By inspection, the relationship between the $\mathbf{q}_i(E)$, $i = 1, 2, \ldots, 7$, and the locus of the Bogoliubov band minima $\mathbf{k}_B(E) = (k_x(E), k_y(E))$ are given by:

$$\mathbf{q}_1 = (2k_x, 0) \tag{1}$$

$$\mathbf{q}_2 = (k_x + k_y, k_y - k_x) \tag{2}$$

$$\mathbf{q}_3 = (k_x + k_y, k_y + k_x) \tag{3}$$

$$\mathbf{q}_4 = (2k_x, 2k_y) \tag{4}$$

$$\mathbf{q}_5 = (0, 2k_y) \tag{5}$$

$$\mathbf{q}_6 = (k_x - k_y, k_y + k_x) \tag{6}$$

$$\mathbf{q}_7 = (k_x - k_y, k_y - k_x) \tag{7}$$

This is referred to[15–19] as the 'octet model'. When these $\mathbf{q}_i(E)$ are measured from the Fourier transform of spatial modulations[16,19] seen in differential tunnelling conductance $dI(\mathbf{r}, V)/dV \equiv g(\mathbf{r}, V)$ the $\mathbf{k}_B(E)$ can be determined by using equations (1)–(7) with the requirement that all independent solutions be consistent. The structure of the superconductor's energy gap $\Delta(\mathbf{k})$ is then determined[15–19] directly from $\mathbf{k}_B(E)$ (see Supplementary Information, sections III and IV). Of primary significance here is that, because only the BQP states of a d-wave superconductor could exhibit such a particle–hole-symmetric set of interference wavevectors in which all dispersions are internally consistent within the octet model, the gap $\Delta(\mathbf{k})$ determined by these procedures is definitely the superconducting energy gap[16,19].

These procedures are demonstrably successful near optimal doping. In $Bi_2Sr_2CaCu_2O_{8+\delta}$, measurements from QPI of $\mathbf{k}_B(E)$ and the superconducting $\Delta(\mathbf{k})$ are consistent with ARPES[16], whereas in $Ca_{2-x}Na_xCuO_2Cl_2$, QPI yields $\mathbf{k}_B(E)$ and $\Delta(\mathbf{k})$ measurements equally well[19]. For unknown reasons, when $E < 3–6$ meV in both systems, the modulations appear to be too weak to analyse (although with unitary scatterers these intra-nodal scattering modulations do appear). In any case, the fundamental $\mathbf{k}$-space phenomenology behind this d-wave BQP interference model[15,17,18] has also been demonstrated directly in ARPES studies[36–38]. QPI studies have been more challenging as $p \to 0$, though, because intense atomic-scale spatial fluctuations in electronic structure cause systematic errors in setting the scanning tunnelling microscope tip elevation[19,20]. Recently, however, it was shown[19] how to enhance the QPI signatures by using the ratio of differential conductances at opposite bias:

$$Z(\mathbf{r}, E = eV) \equiv \frac{g(\mathbf{r}, +V)}{g(\mathbf{r}, -V)}$$

The advantage of this procedure is that it cancels[19,20] the severe systematic errors in $g(\mathbf{r}, V)$ due to tip elevation errors, yet retains all the $\mathbf{q}_i(E)$ information (see Supplementary Information, section II).

**Outline of methods**

Here we report the first application of these techniques to the simultaneous study of **r**-space and **k**-space electronic structure as $p \to 0$ in a copper oxide superconductor. We measure $g(\mathbf{r}, V)$ in ~50 nm × 50 nm fields of view in five $Bi_2Sr_2CaCu_2O_{8+\delta}$ samples, respectively with $p \approx$ 19%, 17%, 14%, 8%, 6% and transition temperatures $T_c$ = 86 K, 88 K, 74 K, 45 K, 20 K. Each sample is inserted into the cryogenic ultrahigh vacuum of the scanning tunnelling microscope and cleaved to reveal an atomically clean BiO surface, and all $g(\mathbf{r}, V)$ measurements are made with atomic resolution and register at or below 4.2 K. This data set consists of >$10^6$ tunnelling spectra and is presented and analysed in detail in the Supplementary Information. From it, the $Z(\mathbf{r}, E)$ and $Z(\mathbf{q}, E)$ are calculated. For all energies at which dispersive $Z(\mathbf{r}, E)$ modulations exist, the octet model is used to find $\mathbf{k}_B(E)$ by means of equations (1)–(7). Because we require comprehensive internal consistency of all $\mathbf{q}_i(E)$ within the octet model, the resulting $\mathbf{k}_B(E)$ is heavily overdetermined (see Supplementary Information, section III, and Supplementary Fig. 2). For QPI analysis at $p = 19\%$, we use the $g(\mathbf{r}, V)$ and $g(\mathbf{q}, V)$ data directly, both because tip elevation errors are minimal and for comparison with previous studies[16].

**Extinction of BQP interference**

In Fig. 2c we show a representative low-energy $Z(\mathbf{q}, E)$ (from a $p$ ~ 8% sample). It reveals the expected 16 pairs of **q** vectors, which are consistent with each other within the octet model. The measured dispersion of each independent **q** vector is shown by open symbols in Fig. 2e. Equivalent data for all hole densities studied are shown in Supplementary Fig. 2. We find that the dispersion of octet model **q** vectors always stops, and that the intensity of all but two **q** vectors diminishes to zero, at some weakly doping-dependent excitation energy which we label $E = \Delta_0$. This is indicated by black arrows in Fig. 2e, f. In Fig. 2d we show a representative $Z(\mathbf{q}, E)$ for $E > \Delta_0$ (from the same $p$ ~ 8% sample). Here we detect only two non-dispersive **q** vectors, labelled $\mathbf{q}_1$* and $q_5^*$ in Fig. 2d, e. The equivalent pair of non-dispersive **q** vectors has been detected using ARPES in optimally doped $Bi_2Sr_2CaCu_2O_{8+\delta}$ (ref. 37) and in underdoped $Ca_{2-x}Na_xCuO_2Cl_2$ (ref. 31). Here, however, we find that this phenomenon occurs in all samples with 6% < $p$ < 19% (the $q_1^*$ and $q_5^*$ data for all $E > \Delta_0$ are shown for these $p$ values in Supplementary Fig. 2). Our first key finding is then that, above some energy $\Delta_0$, the dispersing BQP interference wavevectors $q_i(E)$ always disappear, to be replaced by a completely non-dispersive excitation spectrum represented by $q_1^*$ and $q_5^*$.

In Fig. 3a we plot the $k_B(E)$ determined from the same data. Here it is apparent that when the BQP interference patterns disappear at $\Delta_0$, the **k** states are near the diagonal lines between (0, $\pm\pi/a_0$) and ($\pm\pi/a_0$, 0) within the $CuO_2$ Brillouin zone. This **k**-space 'extinction point' for BQP interference is defined not only by the change from dispersive to non-dispersive characteristics, but also by the disappearance of the $q_2$, $q_3$, $q_6$ and $q_7$ modulations (Fig. 2f and Supplementary Fig. 5). Thus, the BQP interference signatures of delocalized Cooper pairs vanish close to the perimeter of a **k**-space region bounded by lines between (0, $\pm\pi/a_0$) and ($\pm\pi/a_0$, 0). We emphasize that this occurs neither at precisely the same **k** vector nor energy $\Delta_0$ for each $p$, but always near the boundary of this restricted region (Fig. 3a). Within this region, the quasi-particles are confined to a Bogoliubov arc (fine solid lines in Fig. 3a) that shrinks rapidly towards **k** = ($\pm 1/2$, $\pm 1/2$)$\pi/a_0$ with decreasing $p$. We hypothesize that this Bogoliubov arc is always coincident with the Fermi arc detected in the normal state[30–34,36,37].

**Carrier density counts**

Conventional theory would predict that the minima (maxima) of the Bogoliubov bands $k_B(\pm E)$ should occur at the **k**-space location of the Fermi surface of the non-superconducting state. By making this assumption here, we may ask if the carrier density count satisfies Luttinger's theorem, which states that twice the **k**-space area enclosed by the Fermi surface, measured in units of the area of the first Brillouin zone, equals the number of electrons per unit cell, $n$. In Supplementary Fig. 4a we show as fine solid lines hole-like Fermi surfaces fitted to our measured $\mathbf{k}_B(E)$. Using Luttinger's theorem with these $\mathbf{k}$-space contours would result in a calculated hole density $p$ (when measured from half filling, this is defined conventionally as $1 - n$) for comparison with the estimated hole density in the samples. These data are shown by filled symbols in the inset to Fig. 3a. We see that the classic Luttinger theorem is strongly violated at all dopings below $p \sim 10\%$. This is neither a unique nor an anomalous observation: equivalent results have been reported previously for $Ca_{2-x}Na_xCuO_2Cl_2$ at similarly low $p$ values[31].

However, the copper oxides are not metals but carrier-doped Mott insulators. For such systems, Luttinger's theorem must be amended[39] so that the zero-energy contours bounding the $\mathbf{k}$-space region representing carriers are defined not only by poles in the Green's functions, but also by their zeros. Essentially, the perturbation theory description of the metallic Fermi liquid breaks down and correlations among the particles generate zeros in the Green's functions. The locus of zeros of these Green's functions may be expected to occur at the lines joining $\mathbf{k} = (0, \pm\pi/a_0)$ to $\mathbf{k} = (\pm\pi/a_0, 0)$. In that situation, the hole density is related quantitatively to the area between the lines joining $\mathbf{k} = (0, \pm\pi/a_0)$ to $\mathbf{k} = (\pm\pi/a_0, 0)$ and the Fermi arcs. The carrier densities calculated using the region bounded by $\mathbf{k}_B(E)$ (arcs in Fig. 3a) and the hypothesized lines of zeros between $\mathbf{k} = (0, \pm\pi/a_0)$ and $\mathbf{k} = (\pm\pi/a_0, 0)$ (for example the dashed diagonal in Fig. 3a) are shown by open symbols in the inset to Fig. 3a. These are in better agreement with the estimated hole density (see Supplementary Information). Thus, we conclude that if the Green's function lines of zeros occur between $\mathbf{k} = (0, \pm\pi/a_0)$ and $\mathbf{k} = (\pm\pi/a_0, 0)$, the measured $\mathbf{k}$-space structure and the doped-hole density can remain consistent as $p \to 0$.

**Relationship of superconducting energy gap to the pseudogap energy**

In Fig. 3b we plot the doping dependence of the superconducting energy gap $\Delta(\theta_\mathbf{k})$ (see Supplementary Information section IV) in terms of $\theta_\mathbf{k}$ the angle in $\mathbf{k}$-space measured about the $(\pi,\pi)$ point as shown in the inset. We find that $\Delta(\theta_\mathbf{k})$ is always cut off by the Bogoliubov quasi-particle extinction at the boundary of the restricted region. Moreover, the energy $\Delta_0(p)$ where BQP interference disappears and the spatially averaged energy $\overline{\Delta}_0(p)$ at which electronic homogeneity is lost in $Bi_2Sr_2CaCu_2O_{8+\delta}$ (ref. 26) are found to be indistinguishable within their uncertainties. The function $\Delta(\theta_\mathbf{k}) = \Delta_{QPI}[B\cos(2\theta_\mathbf{k}) + (1 - B)\cos(6\theta_\mathbf{k})]$, where $\Delta_{QPI}$ is the theoretical superconducting energy gap maximum at $\theta_\mathbf{k} = 0, \pi/2$, is required to fit the measured $\Delta(\theta_\mathbf{k})$ as shown by the fine solid lines in Fig. 3b. We find that with decreasing $p$, $\Delta_{QPI}$ increases rapidly and $B$ decreases slowly (Fig. 3b). Finally, the maximum energy of the fitted superconducting gap $\Delta_{QPI}$ is always in good quantitative agreement with the spatially averaged pseudogap maxima $\langle\Delta_1\rangle$ as derived from the particle–hole-symmetric peaks in the spectra (Fig. 1b); this relationship is shown in the inset to Fig. 3b.

**r-space structure of pseudogap excitations**

Next we examine the structure of excitations above the extinction energy $\Delta_0$, where no dispersive QPI is detected. We find that these $Z(\mathbf{q}, E)$ have only two non-dispersive $\mathbf{q}$ vectors, namely $\mathbf{q}_1^*$ and $\mathbf{q}_5^*$, which evolve with $p$ as shown implicitly in Fig. 3a (and in detail in Supplementary Fig. 6). As might be expected from their lack of energy dispersion, it is in $\mathbf{r}$ space that these excitations appear most well defined. Analysis of $Z(\mathbf{r}, E)$ for $\Delta_0 < E < 150$ meV shows spatial patterns that are highly similar at all energies but have spatial variations in intensity. Representative examples are shown in Fig. 4a, b. The patterns are short-correlation-length Cu–O–Cu bond-centred modulations in $Z(\mathbf{r}, E)$ with nanoscale unidirectional domains $\sim 4a_0$ wide embedded in a glassy matrix. The spatial structure in these $\mathbf{r}$-space patterns (Fig. 4a, b) appears closely related to that detected by maps of:

$$R(\mathbf{r}, E = eV) \equiv \frac{I(\mathbf{r}, +V)}{I(\mathbf{r}, -V)}$$

These quantify variations in the energy-integrated tunnelling asymmetry, as described in ref. 20 for $V = 150$ mV; their spatial arrangement forms a Cu–O–Cu bond-centred electronic pattern with dispersed ~$4a_0$-wide unidirectional nano-domains. However, because these maps integrate over energy, they do not reveal the characteristic energy of the constituent $\mathbf{r}$-space phenomena.

To address this issue, we focus on the maximum intensity of $Z(\mathbf{r}, E)$ for each $E$. This fluctuates strongly in space as shown, for example, in Fig. 4a, b. However, simultaneous images of the pseudogap energy scale $\Delta_1(\mathbf{r})$ (as defined in the inset to Fig. 4d) also show strong spatial fluctuations (Fig. 4c). Comparing these with Fig. 4a, b, it seems that $Z(\mathbf{r}, E)$ exhibits its maximum intensity in the spatial regions where $E = \Delta_1(\mathbf{r})$. To quantify this, we scale the energy $E$ at each $\mathbf{r}$ by the pseudogap magnitude $\Delta_1(\mathbf{r})$ at the same location, thus defining the new energy scale $e(\mathbf{r}) = E/\Delta_1(\mathbf{r})$ to be a fraction of the local pseudogap energy scale. We find that the translational- and $C_4$-symmetry-breaking bond-centred modulations exhibit an apparent maximum intensity at $e = 1$, or $E(\mathbf{r}) = \Delta_1(\mathbf{r})$ (Supplementary Information section VII and Fig. 7). Our conclusion is then that the intricate $\mathbf{r}$-space patterning of electronic structure seen in the maps of $R$ (ref. 20 and Fig. 4f) is actually an atomic-scale visualization of the spatial structure of low-$p$ pseudogap excitations (Figs 4e and 5a and Supplementary Fig. 7).

**Summary and discussion**

As p is reduced towards the Mott insulator state, scattering interference modulations of BQPs always disappear at an energy $\Delta_0$ that is indistinguishable from the energy at which electronic homogeneity is lost[26]. BQP interference disappears near the perimeter of a region in $\mathbf{k}$ space restricted by the lines joining $\mathbf{k} = (0, \pm\pi/a_0)$ and $\mathbf{k} = (\pm\pi/a_0, 0)$. For energies $E > \Delta_0$, the electronic structure appears to be static in $\mathbf{r}$-space and independent of *E*. In fact, it consists of the atomic-scale spatial patterns previously reported[20] but here identified as the pseudogap excitations at $E = \pm\Delta_1$. Our observations therefore provide a new and different context within which to understand the two excitation energy scales as $p \to 0$. The lower energy, $\Delta_0$, is associated with the disappearance of the BQP interference arising from the presence of delocalized Cooper pairs, whereas the upper energy, $\Delta_1$, is associated with the characteristic **r**-space electronic structure of the pseudogap excitations. Overall, a progressive conversion from the former to the latter electronic structure occurs as *p* decreases to zero even though their characteristic energies $\Delta_{QPI}$ and $\Delta_1$ remain equal. Perhaps most notably, the low-*p* pseudogap excitations locally break the translational symmetry, and reduce the $C_4$ symmetry of the electronic structure in each four-copper-atom plaquette to $C_2$ symmetry in Cu–O–Cu bond-centred patterns without long-range order[20].

A number of theoretical issues emerge from these observations. For example, we do not know why the BQP interference extinction occurs near the perimeter of the k-space region bounded by the lines joining $\mathbf{k} = (0, \pm\pi/a_0)$ and $\mathbf{k} = (\pm\pi/a_0, 0)$. One reason could be that this **k**-space perimeter coincides with the boundary at which elastic particle–particle Umklapp scattering should intensify rapidly as the Mott insulator state is approached[1]. A different possibility is that the **r**-space electronic structure has undergone a reconstruction in which the periodicity of the crystal structure is increased due to the appearance of a coexisting long-range-ordered state; the arcs in Fig. 3a would then represent a hole pocket within a reduced Brillouin zone. However, neither the antiferromagnetism nor the other electronic long-range orders[40,41] necessary for such a reconstruction are observed in $Bi_2Sr_2CaCu_2O_{8+\delta}$. A related explanation could be inelastic scattering of the quasi-particles by spin fluctuations[42,43] at $\mathbf{Q} = (\pi, \pi)$ or by fluctuations of a circulating-current order[40,41] (each would exhibit a reconstruction if stabilized). A key difficulty with all these possibilities is the loss of translational symmetry and local reduction of the expected $C_4$ symmetry of the Cu plaquette to $C_2$

(Fig. 4a, b, e, f and ref. 20), which are not required within such models. Yet another explanation could be static spin–charge stripe glass coexisting with superconductivity[44–47]. This could explain the loss of translational symmetry and the reduction of $C_4$ symmetry to $C_2$ within the pseudogap excitations. However, it does not explain the locus of quasi-particle extinction along the lines joining $\mathbf{k} = (0, \pm\pi/a_0)$ and $\mathbf{k} = (\pm\pi/a_0, 0)$. A related class of proposals, perhaps more congruent with our data, suggest that the fluctuations of an order parameter that would break translational and $C_4$ symmetry if stabilized scatter the quasi-particles strongly[26] and alter the QPI processes[48–50] in a way consistent with our observations.

Notwithstanding these theoretical issues, a new empirical model of the bipartite electronic structure of copper oxides emerges here (Fig. 5). Its key elements include: with falling $p$, the BQP interferences indicating the presence of delocalized Cooper pairs disappear near the lines connecting $(\pm\pi, 0)$ and $(0, \pm\pi)$ in $\mathbf{k}$-space - even as the Cooper pairing energy continues to rise; although the pseudogap excitations appear to have the same energy scale, $\Delta_1$, as the maximum energy scale of superconductivity, $\Delta_{QPI}$, they have a radically different $\mathbf{r}$-space phenomenology that locally breaks the $C_4$ symmetry on Cu plaquettes down to $C_2$ in a bond-centred pattern. To determine the validity of this new model, two immediate implications can be tested. First, $C_4$ symmetry should be recovered on each Cu plaquette at even lower hole density as the antiferromagnetism reappears. Second, the diverse repercussions of this bipartite electronic structure for the bulk thermodynamic characteristics should be compared with the results of relevant measurements.

We acknowledge and thank A. V. Balatsky, J. C. Campuzano, E. Fradkin, A. Georges, T. Hanaguri, P. J. Hirschfeld, S. Kivelson, E.-A. Kim, G. Kotliar, P. A. Lee, M. Norman, P. Phillips, M. Randeria, T. M. Rice, S. Sachdev, K. Shen, Z. X. Shen, A. Tsvelik, M. Vojta and F. C. Zhang for discussions. This work was supported by the US National Science Foundation through the Cornell Center for Material Research, by Brookhaven National Laboratory, by the US Department of Energy, by the US Office of Naval Research, by a Grant-in-Aid for Scientific Research from the Ministry of Science and Education (Japan), and by the 21st Century COE Program of the Japan Society for the Promotion of Science. P.W. acknowledges support from the Humboldt Foundation and A.S. acknowledges support from the US Army Research Office..

Reprints and permissions information is available at www.nature.com/reprints. Correspondence and requests for materials should be addressed to J.C.D. (jcdavis@ccmr.cornell.edu).

**Figure 1 Two classes of electronic excitations in copper oxides as $p \rightarrow 0$.** **a**, The separation between the energy scales associated with excitations of the superconducting state ($E_{SC}$) and those of the pseudogap state ($E_{PG}$) increases as $p$ decreases (reproduced from ref. 22). The different symbols indicate different experimental techniques. **b**, The average of all conductance spectra exhibiting the spatially averaged value of $\Delta_1$, for each of six independently measured $Bi_2Sr_2CaCu_2O_{8+\delta}$ samples. The observed gap maximum increases along the $E_{PG}$ trajectory as hole density is reduced[23]. Note that the peak energies $\pm E_{PG}$ are symmetric about the Fermi level (except perhaps at $p \sim 6\%$ or $T_c \sim 20$ K where they are quite difficult to discern). **c**, The conventional view

20of the electronic structure in one-quarter of the $CuO_2$ first Brillouin zone with energy ($E$) plotted as a third dimension. The d-wave superconducting energy gap $E_{SC}$, which opens for $T < T_c$, is shown in orange, and the temperature-independent pseudogap, which opens for $T \gg T_c$, is shown in blue.

**Figure 2 BQP interference**. **a**, The contours of constant energy in the superconducting state. The locations of the eight regions of maximum joint density of states are shown by coloured symbols. Quasi-particle scattering between these eight regions produces interference patterns associated with the dominant modulations in d$I$(**r**, $V$)/d$V$. The expected wavevectors of these modulations are labelled $q_i$, $i$ = 1, 2, …, 7, in the appropriate colour. **b**, The set of 16 pairs of wavevectors $q_i$ ($E$) representing the octet model described in **a** constitute a highly overdetermined set in **q** space. **c**, The **q**-space structure of interference patterns $Z$(**q**, $E$ = 16 mV) measured on the sample with $T_c$ = 45 K is consistent with the scattering patterns predicted from the octet model ideas of **a** and **b**. **d**, For comparison with **c**, we show $Z$(**r**, $E$ = 38 mV), in which the interference patterns are no longer obviously associated with quasi-particle interference patterns as in **b** and, furthermore, retain this unchanging **q**-space structure for a very wide range of energies $E > \Delta_0$. **e**, The magnitude of various extracted scattering vectors, labelled with the same colours and symbols as in **a** and **b** and plotted as a function of energy. Whereas the expected energy dispersion of the octet vectors $q_i(E)$ is apparent for $|E|$ < 32 mV, the peaks which avoid extinction ($q_1$* and $q_5$*) always become non-dispersive above $\Delta_0$ (black arrow). An interesting precursor effect can be seen in the departure of the dispersive $q_5(E)$ from expectations for $E < \Delta_0$. **f**, The decay of $Z$(**q**, $E$) intensity in all the dispersive modulation wavevectors and its disappearance below the noise level at the extinction energy $\Delta_0$ (black arrow). These effects are described for all data sets in the Supplementary Information.

**Figure 3 Extinction of BQP interference**. **a**, Locus of the Bogoliubov band minimum $k_B(E)$ found from extracted QPI peak locations $q_i(E)$, in five independent $Bi_2Sr_2CaCu_2O_{8+\delta}$ samples with decreasing hole density. Fits to quarter-circles are shown and, as *p* decreases, these curves enclose a progressively smaller area. We find that the BQP interference patterns disappear near the perimeter of a **k**-space region bounded by the lines joining **k** = (0, $\pm\pi/a_0$) and **k** = ($\pm\pi/a_0$, 0). The spectral weights of $q_2$, $q_3$, $q_6$ and $q_7$ vanish at the same place (dashed line; see also Supplementary Fig. 3). Filled symbols in the inset represent the hole count $p = 1 - n$ derived using the simple Luttinger theorem, with the fits to a large, hole-like Fermi surface shown in Supplementary Fig. 4a and indicated schematically here in grey. Open symbols in the inset are the hole counts calculated using the area enclosed by the Bogoliubov arc and the lines joining **k** = (0, $\pm\pi/a_0$) and **k** = ($\pm\pi/a_0$, 0), and are indicated schematically here in blue. **b**, The evolution of the superconducting energy gap $\Delta(k)$ is shown for the **k**-space points shown in **a**, but here as a function of the angle $\theta_k$ about the point ($\pi$, $\pi$) as shown in the lower insert. These are fitted to the parameterization $\Delta(\theta_k) = \Delta_{QPI}[B\cos(2\theta_k) + (1 - B)\cos(6\theta_k)]$ for each hole density (and offset vertically for clarity) as indicated by the fine solid lines. The measured $\Delta(\theta_k)$ values are shown by the coloured data symbols on the left-hand side, and the error bars representing the standard deviations for each measurement are shown on the right-hand side. As *p* decreases, it is obvious that the fitted $\Delta_{QPI}$ increases rapidly. The upper inset shows the relationship between $\Delta_{QPI}$ and the average pseudogap energy $\langle\Delta_1\rangle$.

**Figure 4 Imaging copper oxide pseudogap excitations as $p \to 0$**. **a**, **b**, Atomically resolved $Z(r, E)$ for the $T_c$ = 45 K sample (simultaneous topographic image shown in **d** for two distinct energies (60, 120 mV). The intensity of electronic structure patterns in **r** space vary as a function of energy. **c**, The corresponding spatial map of the $\Delta_1$ gap magnitude over the sample surface in **d**. There is a wide distribution of heterogeneous pseudogap values[26]. **d**, A topograph showing the locations of Bi atoms (small, bright circles) on the field of view where all data in this figure were acquired. The inset shows the local definition of $\Delta_1(\mathbf{r})$ with the tunnelling conductances at $E=\pm\Delta_1$ represented by $g^+$ and $g^-$ respectively. **e**, Image of $Z(\mathbf{r}, e = E/\Delta_1 = 1)$ in which energy has been rescaled by the local value of $\Delta_1(\mathbf{r})$ from **c**; this represents an image of what pseudogap states would look like in terms of $Z(\mathbf{r}, E = \Delta_1)$ if the nanoscale disorder in $\Delta_1$ were not to exist. **f**, The $R(\mathbf{r}, V = 150$ mV$)$ patterns are virtually identical to those in **e**. Thus, the spatial patterning reported in ref. 20 is actually concentrated on the states at $E = \pm\Delta_1$, meaning that these **r**-space excited states are the copper oxide pseudogap excitations as $p \to 0$.

**Figure 5 Bipartite electronic structure of copper oxides as $p \to 0$. a**, The pseudogap excitations are separated from the Bogoliubov quasi-particle states by a plane in $E$–**k** space restricted approximately by the lines joining $\mathbf{k} = (0, \pm\pi/a_0)$ and $\mathbf{k} = (\pm\pi/a_0, 0)$. This is indicated in blue. The energy $E = \Delta_0$ at which the excitations make the transition from being dispersive BQP to being energy independent pseudogap states (and at which spatial homogeneity is lost – see ref. 26), diminishes slowly with decreasing $p$. **b**, For energies $E > \Delta_0$ the electronic structure is static in **r**-space and independent of $E$; the pseudogap states at $|E| = \Delta_1$ are short-correlation-length Cu–O–Cu bond-centred modulations in $Z(\mathbf{r}, E)$ that break translational symmetry and reduce $C_4$ symmetry to $C_2$ within each four-Cu-atom plaquette (see ref. 20). **c**, In contrast, scattering interference between **k**-space BQP excitations from delocalized Cooper pairs is observed on an arc terminated approximately by the plane in $E$-**k** space (blue in **a**) that connects $\mathbf{k} = (0, \pi/a_0)$ and $\mathbf{k} = (\pi/a_0, 0)$. As $p \to 0$, the arc shrinks and the phenomena in **c** rapidly disappear to be replaced by those in **b**.

**Fig. 1**

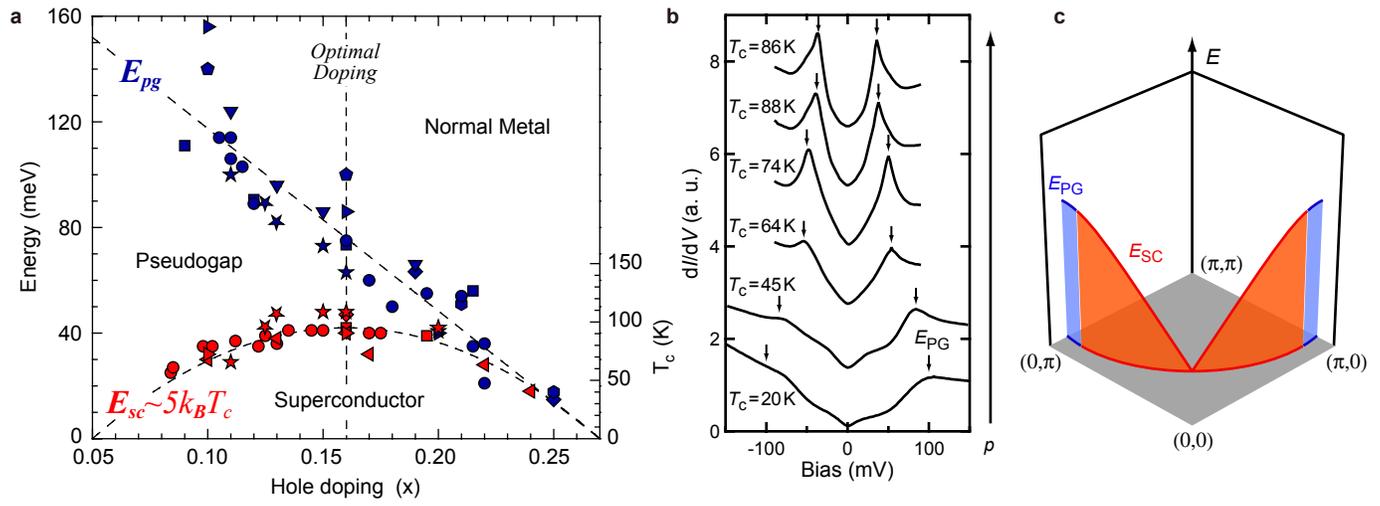

**Fig. 2**

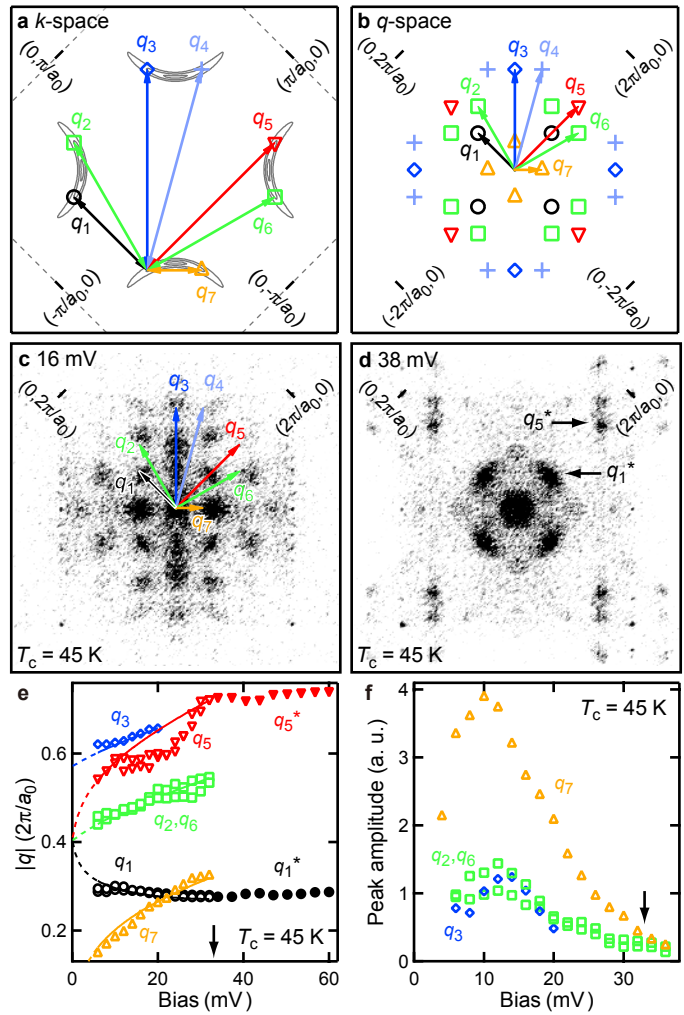

**Fig. 3**

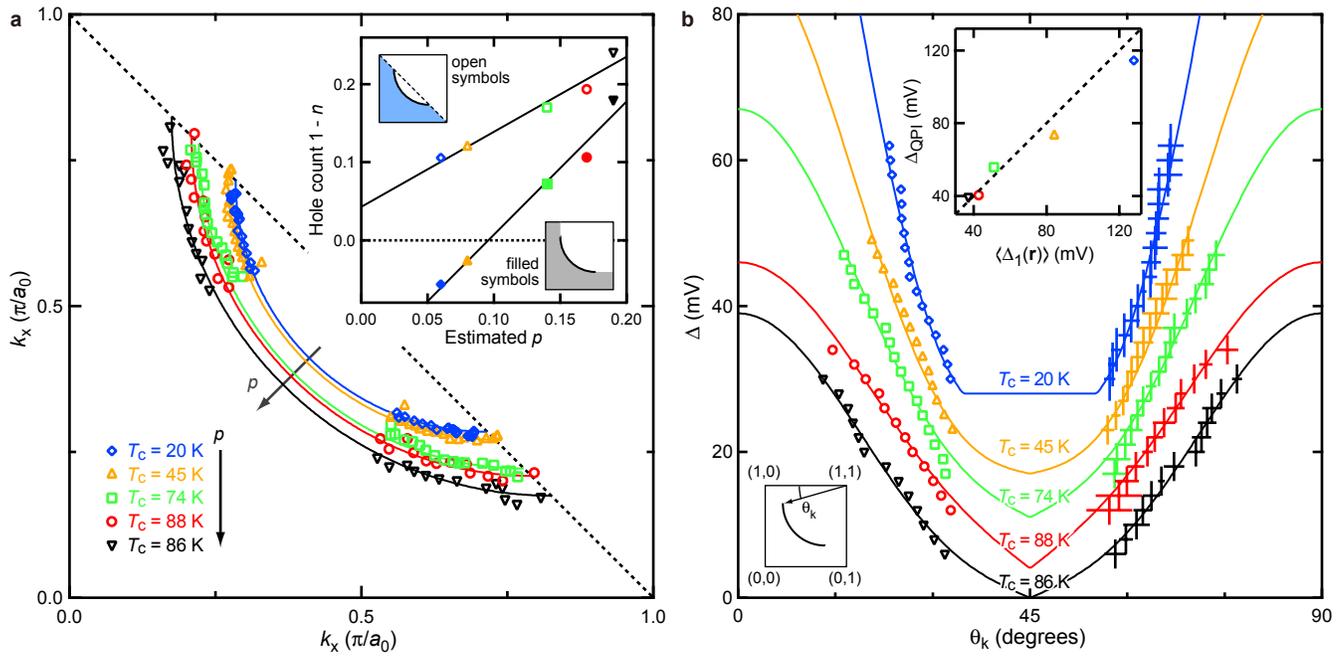

**Fig. 4**

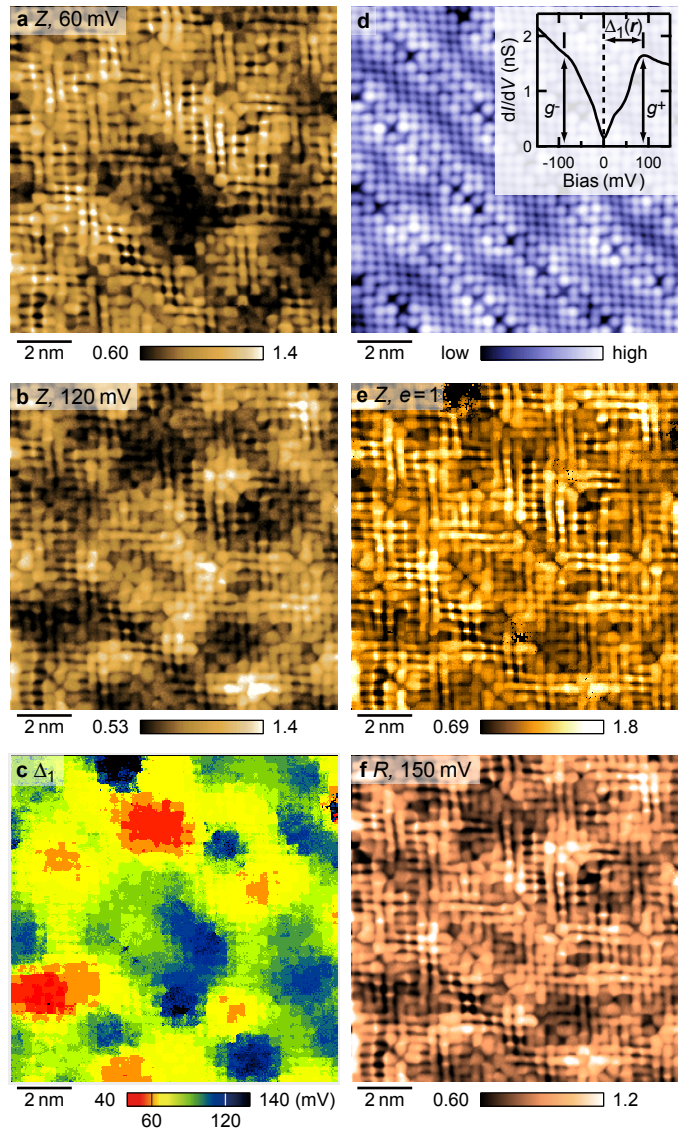

**Fig. 5**

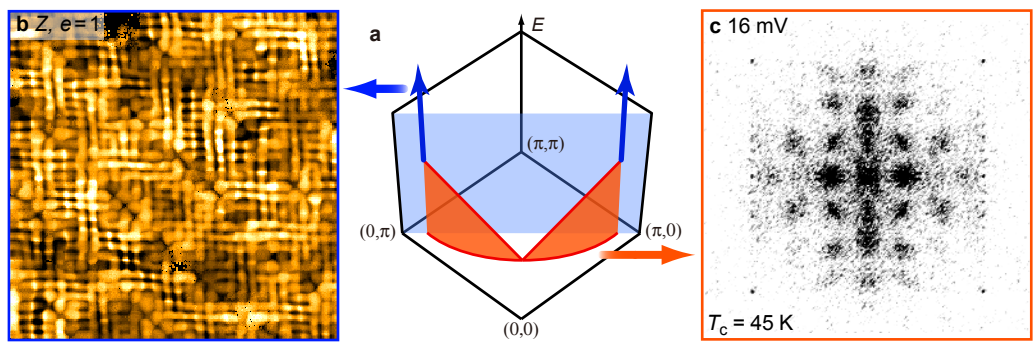